\documentclass[CTAC,short]{cta-v2}

\usepackage{nicefrac}												
\usepackage{amssymb}										
\usepackage{amsmath}	
\usepackage{pdfpages}

\usepackage{float}													
\usepackage{lineno}												

\usepackage{multirow}											
\usepackage{subcaption}											
\usepackage{textcomp}											
\usepackage{array}													
\usepackage{url}
\usepackage{pdflscape}											
\usepackage[backgroundcolor = red]{todonotes}		
\usepackage{textgreek}											
\usepackage{color, soul}											
\usepackage{listings}												

\usepackage{longtable}

\usepackage[numbers]{natbib}
\usepackage[normalem]{ulem}
\usepackage{graphics,graphicx}

\usepackage{color}

\newcolumntype{H}{>{\columncolor{ctablue}}c}
\newcolumntype{S}{>{\columncolor{lightgray}}c}
\newcommand{\sectionm}[1]{{\normalfont\Large\bfseries\ctablue{#1}}}

\begin{document}


\title{Multi-messenger and transient astrophysics with the Cherenkov Telescope Array}
\shorttitle{Multi-messenger and transient astrophysics with CTA}
\editor{THIS IS A TEST, M Astronomer 2}					
\approver{}
\releaser{}
\docnum{}
\issue{0}
\revision{1}
\comment{Draft produced by }
\ownership{\acrshort{CTAO}}
\distribution{\acrshort{PC}}

\history
{
	1 & b & 2019-11-05 & Second draft &  \\ 	
	\hline 	
	1 & a & 2018-12-17 & First draft &  \\ 		
	\hline 				
}

\contributors
{
	\hline 	
	\hline 					
}



\include{definitions}
\include{acronyms}
\include{glos} 
\include{glos_science}





# Multi-messenger and transient astrophysics with the Cherenkov Telescope Array

**Authors:** Ž. Bošnjak (Zagreb University, Croatia), A. M. Brown (Durham University), A. Carosi* (University of Geneva - DPNC, Switzerland), M. Chernyakova (School of Physical Sciences and CfAR, Dublin City University, Ireland), P. Cristofari (Observatoire de Paris, France), F. Longo (University of Trieste and INFN, Trieste, Italy), A. López-Oramas* (IAC and Universidad de La Laguna, Spain), M. Santander* (University of Alabama, USA), K. Satalecka (DESY-Zeuthen, Germany), F. Schüssler (IRFU / CEA Paris-Saclay, France), O. Sergijenko (Taras Shevchenko National University of Kyiv, Ukraine), A. Stamerra (INAF-Osservatorio Astronomico di Roma, Italy), (continues on next page)

**Contact points:** alessandro.carosi@unige.ch, alicia.lopez@iac.es, jmsantander@ua.edu

https://www.cta-observatory.org/about/cta-consortium/

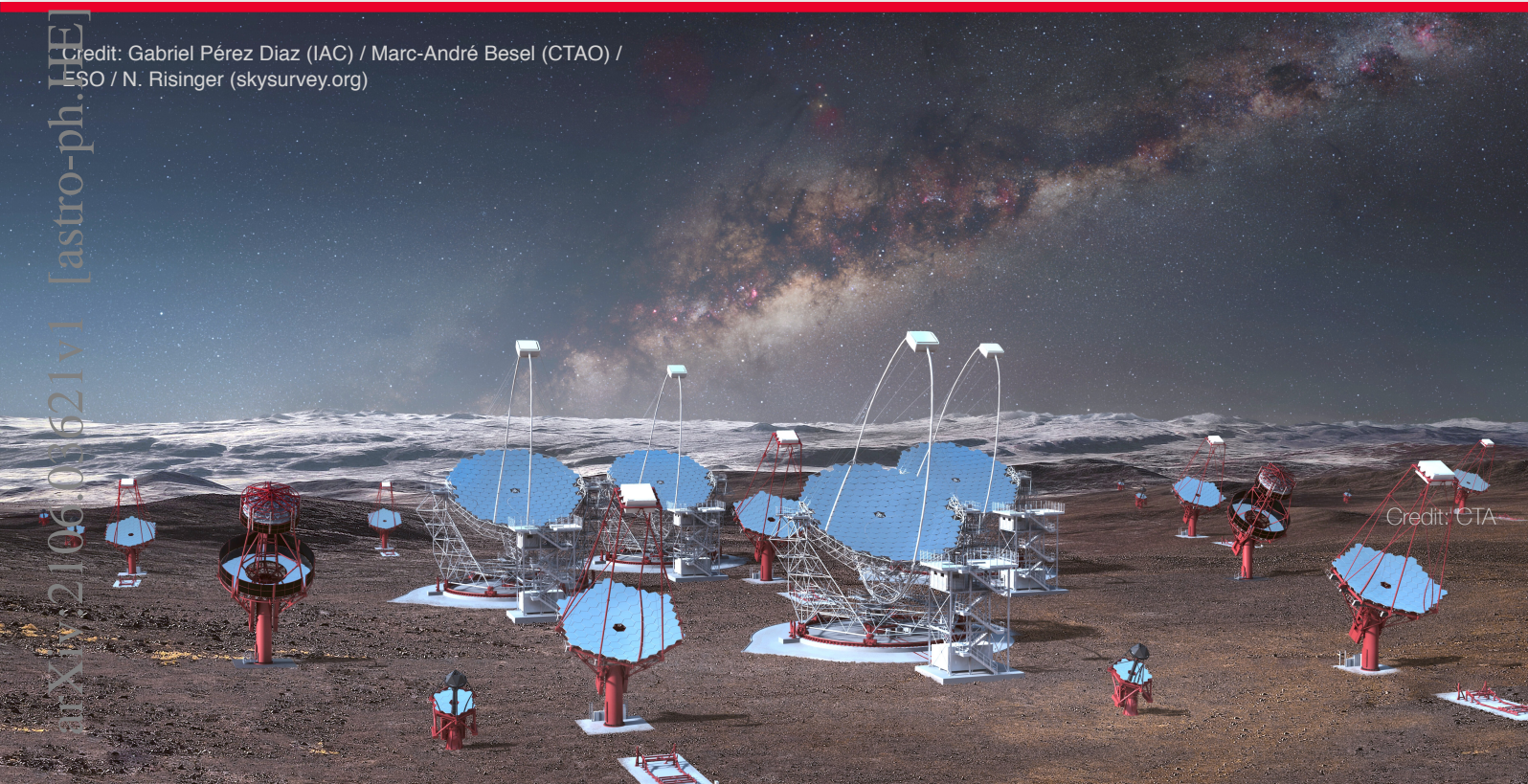

Credit: Gabriel Pérez Diaz (IAC) / Marc-André Besel (CTAO) / ESO / N. Risinger (skysurvey.org)

The discovery of gravitational waves, high-energy neutrinos or the very-high-energy counterpart of gamma-ray bursts has revolutionized the high-energy and transient astrophysics community. The development of new instruments and analysis techniques will allow the discovery and/or follow-up of new transient sources. We describe the prospects for the Cherenkov Telescope Array (CTA), the next-generation ground-based gamma-ray observatory, for multi-messenger and transient astrophysics in the decade ahead. CTA will explore the most extreme environments via very-high-energy observations of compact objects, stellar collapse events, mergers and cosmic-ray accelerators.

# Multi-messenger and transient astrophysics with the Cherenkov Telescope Array

**Authors (continuation):** I. Agudo (Instituto de Astrofísica de Andalucía-CSIC, Spain), R. Alves Batista (Radboud University Nijmegen, the Netherlands), E. Amato (INAF-Osservatorio Astrofisico di Arcetri, Italy), E. O. Anguner (Centre de Physique des Particules de Marseille, France), L. A. Antonelli (INAF-Osservatorio Astronomico di Roma, Italy), M. Backes (University of Namibia, Namibia), Csaba Balazs (Monash University, Australia), L. Baroncelli (INAF-OAS Bologna, Italy), J. Becker Tjus (Ruhr-Universität Bochum, Germany), C. Bigongiari (INAF-Osservatorio Astronomico di Roma, Italy), E. Bissaldi (Politecnico and INFN Bari, Italy), C. Boisson (LUTH, Observatoire de Paris, France), J. Bolmont (LPNHE, CNRS/IN2P3, Sorbonne University, France), M. Böttcher (North-West University, South Africa), P. Bordas (Universitat de Barcelona, ICCUB, IEEC-UB, Spain), C. Braiding (University of New South Wales, Australia), J. Bregeon (LPSC, Univ. Grenoble-Alpes, CNRS/IN2P3), N. Bucciantini (INAF-Osservatorio Astrofisico di Arcetri, Italy), A. Bulgarelli (INAF-OAS Bologna, Italy), M. Burton (Armagh Observatory and Planetarium, UK), F. Cangemi (LPNHE, CNRS, Sorbonne University, France), P. Caraveo (INAF-IASF), M. Cardillo (INAF - IAPS, Roma, Italy),S. Caroff (LAPP, Univ. SMB, CNRS/IN2P3), S. Casanova (IFJ PAN, Krakow, Poland), S. Chaty (University of Paris and CEA Paris-Saclay, France), J. G. Coelho (Universidade Federal do Parana, Brazil), G. Cotter (University of Oxford, UK), A. D'Aì (INAF - IASF Palermo, Italy), F. D'Ammando (INAF-Istituto di Radioastronomia di Bologna, Italy), E. M. de Gouveia Dal Pino (IAG - University of São Paulo, Brazil), D. della Volpe (University of Geneva - DPNC, Switzerland), D. de Martino (INAF-Osservatorio Astronomico di Capodimonte-Napoli, Italy), T. Di Girolamo (Università di Napoli "Federico II", Italy), A. Di Piano (INAF-OAS Bologna, Italy), A. Djannati-Ataï (Université de Paris, CNRS/IN2P3, APC, France), V. Dwarkadas (University of Chicago, USA), E. de Ona Wilhelmi (DESY-Zeuthen, Germany), R. C. Dos Anjos (Universidade Federal do Paraná, Brazil), G. Emery (University of Geneva - DPNC, Switzerland), E. Fedorova (Taras Shevchenko National University of Kyiv), S. Fegan (LLR/Ecole Polytechnique, CNRS/IN2P3), A. Fiasson (LAPP, Univ. SMB, CNRS/IN2P3), V. Fioretti (INAF-OAS Bologna, Italy), M.D. Filipovic (Western Sydney University, Australia), D. Gaggero (Instituto de Física Teórica UAM/CSIC), G. Galanti (INAF - IASF Milano, Italy), D. Gasparrini (INFN Roma Tor Vergata & ASI-SSDC, Italy), G.Ghirlanda (INAF-Osservatorio Astronomico di Brera, Italy), P. Goldoni (APC/IRFU - France) J. Granot (Open University of Israel), J. G. Green (INAF-Osservatorio Astronomico di Roma, Italy), M. (DPNC, Université de Genève), B. Hnatyk (Taras Shevchenko National University of Kyiv, Ukraine), R. Hnatyk (Taras Shevchenko National University of Kyiv, Ukraine), M. Heller (University of Geneva - DPNC, Switzerland), D. Horan (LLR/Ecole Polytechnique, CNRS/IN2P3, France), T. Hovatta (FINCA, University of Turku, Finland), S. Inoue (RIKEN, Japan), M. Jamrozy (Jagielonian University, Poland), D. Kantzas (API/GRAPPA, University of Amsterdam, the Netherlands), B. Khélifi (APC, IN2P3/CNRS- Université de Paris, France), N. Komin (University of the Witwatersrand, Johannesburg, South Africa), A. Lamastra (INAF-Osservatorio Astronomico di Roma, Italy), N. La Palombara (INAF - IASF Milano), J. P. Lenain (LPNHE, CNRS/IN2P3, Sorbonne, France), E. Lindfors (FINCA, University of Turku, Finland), I. Liodakis (FINCA, University of Turku, Finland), S. Lombardi (INAF-Osservatorio Astronomico di Roma, Italy), F. Lucarelli (INAF-Osservatorio Astronomico di Roma and ASI-SSDC, Italy), P. L. Luque-Escamilla (University of Jaen, Spain), P. Majumdar (Saha Institute of Nuclear Physics, Kolkata, India), A. Marcowith (Laboratoire Univers et Particules de Montpellier, France), S. Markoff (API/GRAPPA, University of Amsterdam, the Netherlands), J. Marti (University of Jaen, Spain), M. Martinez (Institut de Física d'Altes Energies, IFAE-BIST, Barcelona, Spain), D. Mazin (ICRR, University of Tokyo, Japan and MPI for physics, Germany), S. McKeague (School of Physical Sciences and CfAR, Dublin City University, Ireland), S. Mereghetti (INAF - IASF Milano, Italy), E. Mestre (ICE-CSIC, Spain), T. Montaruli (University of Geneva - DPNC, Switzerland), G. Morlino (INAF-Osservatorio Astrofisico di Arcetri, Italy), A. Morselli (INFN Roma Tor Vergata), C. Mundell (University of Bath, UK), T. Murach (DESY Zeuthen, Germany), L. Nava (INAF-Osservatorio Astronomico di Brera, Italy), A. Nayerhoda (Institute of Nuclear Physics Polish Academy of Sciences, Krakow, Poland), L. Nicastro (INAF-OAS Bologna, Italy), J. Niemiec (Institute of Nuclear Physics Polish Academy of Sciences, Krakow, Poland),  M. Nikolajuk (University of Bialystok, Poland ), B. Olmi (INAF - Osservatorio Astronomico di Palermo, Italy), R. Ong (University of California, Los Angeles, CA, USA), M. Orienti (INAF - Istituto di Radioastronomia di Bologna, Italy), J.P. Osborne (University of Leicester, UK), Josep M. Paredes (Universitat de Barcelona, ICCUB, IEEC-UB, Spain), G. Pareschi (INAF-Osservatorio Astronomico di Brera, Italy), N. Parmiggiani (INAF-OAS Bologna, Italy), B. Patricelli (INAF - Osservatorio Astronomico di Roma), A. Pe'er  (Bar Ilan University, Israel and University College Cork, Ireland), G. Piano (INAF-IAPS, Roma, Italy), G. Pühlhofer (IAAT, University of Tübingen, Germany), M. Punch (APC, IN2P3/CNRS - Université de Paris, France), O. Reimer (Innsbruck University, Institute for Astro- and Particle Physics, Austria), M. Ribó (Universitat de Barcelona, ICCUB, IEEC-UB, Spain), G. Rodriguez (INAF - IAPS, Roma, Italy), J. Rodriguez (AIM, CEA, CNRS, Université Paris-Saclay, Université de Paris, France),  P. Romano (INAF-Osservatorio Astronomico di Brera, Italy), G. Romeo (INAF - Osservatorio Astrofisico di Catania, Italy), M. Roncadelli (INFN - Pavia, Italy), G. Rowell (University of Adelaide, Australia), B. Rudak (CAMK PAN, Torun, Poland), I. Sadeh (DESY-Zeuthen, Germany), F. Salesa Greus (Institute of Nuclear Physics Polish Academy of Sciences, Krakow, Poland), F. G. Saturni (INAF-Osservatorio Astronomico di Roma, Italy), U. Sawangwit (National Astronomical Research Institute of Thailand, Thailand), M. Seglar-Arroyo (LAPP, Univ. SMB, CNRS/IN2P3), R. C. Shellard (CBPF-Centro Brasileiro de Pesquisas Físicas, Brazil), H. Sol (LUTH, Observatoire de Paris, CNRS, France), R. Starling (University of Leicester, UK), T. Stolarczyk (AIM, CEA, CNRS, Université Paris-Saclay, Université de Paris, France), G. Tagliaferri (INAF-Osservatorio Astronomico di Brera, Italy), F. Tavecchio (INAF-Osservatorio Astronomico di Brera, Italy), L. Tibaldo (IRAP, Université de Toulouse, CNRS, UPS, CNES, France), V. Testa (INAF-Osservatorio Astronomico di Roma, Italy), S. Vercellone  (INAF-Osservatorio Astronomico di Brera), A. Viana (IFSC, Universidade de São Paulo,  Brazil), J. Vink (API/GRAPPA, University of Amsterdam, Netherlands), V. Vitale  (INFN Roma Tor Vergata, Italy), S. D. Vergani (GEPI-Observatoire de Paris, PSL University, CNRS, France), S. Vorobiov  (University of Nova Gorica, Slovenia), A. Wierzcholska (Institute of Nuclear Physics Polish Academy of Sciences, Krakow, Poland),L. Yang  (Sun Yat-sen University, China), D. Zavrtanik (University of Nova Gorica, Slovenia), V. Zhdanov (Taras Shevchenko National University of Kyiv, Ukraine) **on behalf of the CTA Consortium**



\sectionm{Introduction}


The study of very-high-energy (VHE, E$>$100 GeV) gamma rays sits at the interface of the two major emerging fields: multi-messenger and time-domain astrophysics. The connection rests on the fact that these photons can be produced in extreme environments associated with cosmic-ray acceleration (linked with the production of high-energy neutrinos), the relativistic bulk motion of matter in stellar collapses and compact object mergers (which powers supernovae and gravitational wave emission and is also linked to gamma-ray bursts), and the dynamic processes common in many energetic transient and highly-variable emitters such as magnetars, microquasars, and pulsar wind nebulae.

The current generation of imaging atmospheric Cherenkov telescopes (IACTs: H.E.S.S.~\cite{Aharonian:2006pe}, MAGIC~\cite{2016APh....72...76A}, VERITAS~\cite{Holder:2006px}) has already started to explore this field achieving major discoveries, including the first detection of gamma-ray bursts (GRBs) in the VHE range~\cite{MAGIC_GRB190114C_1, HESS_GRB180720B, HESS_GRB190829A} and the discovery and characterization of the first candidate electromagnetic counterpart to a high-energy neutrino detected by IceCube~\cite{Ahnen:2018mvi, VERITAS_TXS}. In the coming decade the premier facility for VHE astrophysics will be the Cherenkov Telescope Array (CTA) observatory, which will consist of two IACT arrays (one in the northern and one in the southern hemisphere), providing all-sky access in the energy range from 20 GeV to more than 100 TeV. CTA will enable unique studies of the multi-messenger and transient sky, as it will operate simultaneously with facilities such as the LIGO/Virgo/KAGRA gravitational wave observatories; the IceCube-Gen2, KM3NeT and GVD neutrino telescopes; the Vera Rubin Observatory; and a large number of next-generation high-energy instruments dedicated to the detection and study of transients and variable objects such as THESEUS and SVOM. We here present an overview of the capabilities of CTA to perform studies of astrophysical transients and multi-messenger triggers, which is one of nine Key Science Projects for the observatory~\cite{Acharya2019}. 



\chapter{Gamma-ray bursts}
\label{sec:GRB}
Gamma-ray bursts are brief and intense pulses of mainly sub-MeV gamma rays releasing as much as $\sim 10^{51}-10^{55}$ ergs of isotropic equivalent energy. They represent the most electromagnetically luminous events in the Universe. A widely accepted interpretation~\cite[see][for a general review]{kumarzhang2015} is that GRB emission arises in a dissipation process in which the energy of the relativistic jet - produced following the collapse of a very massive star or the merger of binary neutron stars or a neutron star and a black hole - is reconverted into non-thermal radiation. Several mechanisms have been suggested for this dissipation, such as internal shocks or magnetic reconnection if the outflow is highly magnetized. Within this framework, synchrotron radiation from high-energy electrons has been considered as the most natural mechanism to explain the observed GRB sub-MeV emission \cite{2018A&A...616A.138O}. Although it cannot fully explain the observed spectral properties for the majority of the events, synchrotron emission is believed to play an essential role in the observed emission \cite{daigne2011} being also largely used as explanation of GRBs observed in the $\sim$few GeV energy band by {\it Fermi}-LAT~\cite[e.g.][]{2018IJMPD..2742003N}. The GeV spectra observed by LAT are mostly consistent with extrapolations of the sub-MeV spectrum to higher energies with no apparent cut offs. 

In particular, it has been suggested that the high-energy (HE, E$>$100 MeV) emission observed by LAT extending after the end of the prompt emission is afterglow-synchrotron radiation produced in the external shock that is driven by a jet into the circum-burst medium. However, observations above few tens of GeV could severely challenge this scenario as the synchroton burn-off limit is exceeded. Therefore, the recent IACT detections of VHE gamma rays from GRB~180720B~\cite{HESS_GRB180720B}, GRB~190114C~\cite{MAGIC_GRB190114C_1}, GRB~190829A~\cite{HESS_GRB190829A} and GRB~201216C~\cite{2020GCN.29075....1B} have represented an important step-forward in our comprehension of GRB physics. The detailed multi-wavelength study of GRB 190114C has shown convincingly and for the first time how the spectrum from X-ray to TeV energies requires an extra spectral component to explain the flux increase at the highest energies~\cite{MAGIC_GRB190114C_1}. Because of the analogies between the temporal profiles of the TeV, GeV and X-ray components, this VHE emission is believed to originate in the same physical regions as the lower-energy radiation. Consequently, it has been proposed that the extra component is generated by synchrotron photons Compton up-scattered by the same electrons accelerated in the shock, i.e. a synchrotron and synchrotron self-Compton (SSC) emission scenario commonly invoked to explain the broadband continuum emission from blazars. However, the phenomenology observed in the current sample of VHE GRBs reveals some challenging aspects: in the case of GRB~180720B, although the observed TeV signal is well above the synchrotron burn-off limit confirming, as for GRB~190114C, the presence of an additional emission component, VHE emission is detected much deeper in the afterglow phase up to $T_0+10$~h. As the flux and maximum energy of the afterglow emission decrease over time, due to the deceleration of the outflow, the detection of a VHE signal at these times challenges current emission scenarios and motivates the GRB follow-up with better sensitivity above $\sim$10 GeV as promised by CTA. This way, it will be possible to explore under which conditions the VHE signal can emerge from the prompt-to-afterglow phase (as for GRB~190114C), or even at a later time. The detection prospects for such CTA observations are necessarily still preliminary and dependent on the final array configuration and performances. However, a detection rate of the order of a few bursts per year might be expected \cite{inoue2013}. Once a sufficiently large number of events is collected, a detailed VHE population study will be possible allowing us to extend the accessible GRB physical parameter space and to understand whether a VHE component is a common feature in GRBs. Furthermore, the achievable photon statistics would allow CTA to better study the spectral and temporal properties of the VHE emission, distinguishing between the variable prompt GRB emission and temporally smoother afterglow emission, and in turn shed light into unresolved issues like determining the dynamics of jet formation and the mechanisms of particle acceleration. In addition, once a robust and reliable comprehension of their intrinsic properties is achieved, CTA will use GRBs as probes of cosmic-ray physics, observational cosmology and fundamental physics.

The high sensitivity and fast reaction time of CTA, combined with multi-wavelength observations, will allow CTA to expand the VHE horizon with respect to the one of the current generation of IACTs and will thus benefit from the increased number of high-redshift GRBs detected in the 2030s by next generation instruments like SVOM and THESEUS~\cite{2020ChA&A..44..269Y, 2018AdSpR..62..191A}. Synergies with future facilities will indeed play a major role in the near future also for the exploration of peculiar events such as the study of short GRBs associated with gravitational wave events and their possible kilonova emission. These observations will be essential to shed light over some still-unanswered questions such as the origin of heavy elements and the chemical evolution of the Universe.

\chapter{Gravitational Waves }
\label{sec:GW}
The era of gravitational wave (GW) astronomy began with the first direct detection of GWs from the coalescence of a binary system composed of two stellar-mass black holes (BH), followed by the multi-messenger observation of the coalescence of two neutron stars (NS, GW170817, ~\citep{2017PhRvL.119p1101A}) \comment{AS: these references can be removed if not needed} and the associated electromagnetic emission in the form of a GRB and a kilonova.
The contemporaneous electromagnetic and GW detection did reveal the richness of information and the complementarity of the two probes, providing a deeper knowledge and details of the merger of the two neutron stars.
 Gravitational waves encode important information (e.g. mass distribution, geometry, etc.) about the system even before it triggers the explosions detected as short GRBs in the electromagnetic domain. Combined observations of GWs, X-rays and VHE gamma rays will thus help to significantly increase our understanding on how the progenitor system links to the observational appearance of the GRB. 
 
 Over the last years the the search for
 VHE emission from GW170817~\cite{2017ApJ...850L..22A, 2020ApJ...894L..16A}  by H.E.S.S. and the hint for emission from the short GRB 160821B by MAGIC~\cite{2020arXiv201207193M} provided first indications on the scientific potential of these searches. Thanks to its superior sensitivity and the increased number of events, CTA will be able to increase the number of detections and to provide a deeper insight on their physical processes. The detection of GeV-TeV emission, as a second high-energy component (as envisaged in GRB190114C, see Sec.~\ref{sec:GRB}) or as single X-ray to TeV component, are key to assess the conditions of the burst progenitor, the parameters of the burst, and the environment necessary to trigger VHE emission, as well as determine the maximum energy reached, its possible scaling with the jet opening angle, and in general to reduce the uncertainties in the non-thermal emission from present GRB models. 

The fast and reliable identification of the EM counterpart depends greatly on the uncertainty in the GW sky localization, spanning an area from a few tens up to hundreds of square degrees.The array of telescopes of CTA can be proficiently used to cover the full GW region in a short time, implementing an optimal strategy (e.g. as in \citep{2021JCAP...03..045A}). The superior short-time sensitivity of the Cherenkov telescopes (see \cite{CTASensitivity}) will potentially lead to the detection of a gamma-ray counterpart that is uniquely associated to the GW event, thanks to the small contamination by gamma-ray sources. 
The rapid reaction, high sensitivity, and realtime analysis of its data stream will make CTA an important contributor to this global effort. Alerts on the detection of VHE counterparts will become promptly available to the community to guide further observations, as soon as 30 s after the data has been recorded. While preliminary studies have been performed in ~\citep{2018JCAP...05..056P}, the CTA GW follow-up program is currently being defined and implemented \citep{2019ICRC...36..790S}.


\chapter{High-energy neutrinos  }

High-energy neutrinos and gamma rays should be jointly produced in astrophysical sources through the hadronic interactions of cosmic rays (CR), charged particles such as protons and atomic nuclei that reach Earth with energies up to $\sim10^{21}$ eV.
While the origin of CRs remains largely unknown, sources such as AGN, star-forming galaxies, supernova remnants, or GRBs that are also VHE gamma-ray emitters are among the leading CR source candidates. The VHE gamma rays from these objects can be produced leptonically, but a coincident observation of neutrinos would clearly identify them as CR accelerators.
%
The observation of astrophysical neutrinos in the TeV-PeV range by IceCube~\cite{Aartsen:2013jdh, 2020arXiv201103545A}, followed by the identification of first evidence for a candidate neutrino source due to the correlated observation of a high-energy neutrino from the direction of the flaring gamma-ray blazar TXS 0506+056~\cite{IceCube:2018dnn} are important steps forward in the search for neutrino emission from gamma-ray sources.

\begin{figure}[h!]
    \centering
    \includegraphics[width=0.5\textwidth]{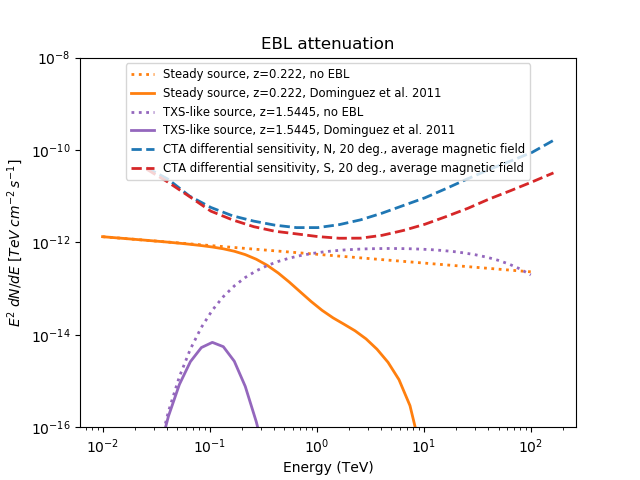}\includegraphics[width=0.5\textwidth]{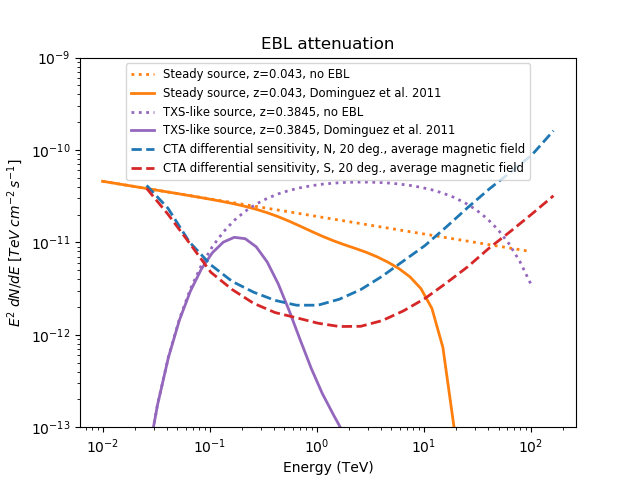}
    \caption{The energy spectra of potential neutrino sources (without and with the EBL attenuation) overlaid on the CTA differential sensitivity for the undetected (left) and detected (right) sources.}
    \label{fig:neutrinoebl}
\end{figure}

 While current-generation IACTs operate active neutrino follow-up programs \cite{Aartsen:2016qbu,2017ICRC...35..618S, IceCube:2018dnn,2019ICRC...36..787S,2019ICRC...36..633B,2019ICRC...36..782S} the sensitivity of these studies is limited as VHE gamma rays may be absorbed or down-scattered while they escape their source or during propagating over cosmological distances due to the effect of the extragalactic background light (EBL, \cite{Hauser:2001xs, 2013APh....43..112D}). CTA, with its fast reaction time and lower energy threshold, will enable sensitive searches for VHE counterparts to well-localized, likely astrophysical neutrino events~\cite{Blaufuss:2019fgv} up to much higher redshifts~(Fig. \ref{fig:neutrinoebl}) enabling counterpart detections or the placing of strong constraints on source opacities. 
 


Recent studies~\cite{Satalecka:2020bao} indicate that for neutrino flares from blazars, CTA will detect a counterpart for about one third of the cases after only 10 mins of observations, with lower detection probabilities for steady neutrino sources. These detections could be performed through dedicated follow up observations of neutrino alerts, or while CTA performs its extragalactic sky survey or the long-term monitoring of flaring AGN. The study of sources that could be linked with neutrino emission, such as tidal disruption events~\cite{Stein:2020xhk}, or that are potential hadronic emitters (specially those extending to PeV gamma-ray energies in our galaxy) will also benefit from joint neutrino studies involving CTA. The sensitivity of these multi-messenger observations will be greatly enhanced by the operation of next generation neutrino telescopes (IceCube-Gen2~\cite{IceCubeGen2}, KM3NeT~\cite{Adrian-Martinez:2016fdl}, Baikal-GVD~\cite{Safronov:2020dtw} and P-ONE~\cite{Agostini:2020aar}).

\chapter{Galactic transients }

A wide range of sources in our Galaxy exhibit transient emission via accretion/ejection processes and interactions between, e.g. jets, outflows and/or strong winds. These events can accelerate particles up to relativistic energies, leading to the production of high-energy radiation. Some of these Galactic transient sources include flares from pulsar wind nebulae (PWNe), where relativistic outflows are driven by the energy loss of rotating neutron stars (NS), jet ejection in microquasars (where a compact object accretes from a companion star) or outbursts and flares from magnetars (NS with high magnetic fields) which have recently been associated with Fast Radio Bursts (FRBs) for the first time \cite{Andersen2020}. Some objects such as novae (explosions from the surfaces of white dwarf stars) or transitional millisecond pulsars (tMSPs, pulsars which change from an accretion to a radio loud phase) have already been detected in the MeV-few GeV regime \cite{FermiNova,ray2012radio}, including the recent discovery of GeV emission from an extragalactic magnetar giant flare \cite{Fermi2021}, but many of these Galactic transient sources have never been detected at VHE. Although certain objects show persistent emission and/or periodically emit variable radiation, we focus on those emitters which display  unpredictable and irregular transient emission at different wavelengths. \par 

CTA, with its unprecedented sensitivity to short-timescale transient events, will likely, for the first time, detect the most energetic counterparts of several Galactic transient sources, unveiling the production and acceleration mechanisms at work. CTA observations of these objects will be based on external triggers from monitoring instruments, such as X-ray or HE satellites. However, the serendipitous discovery of new Galactic transient sources while performing observations, for instance, during the Galactic Plane Survey (GPS), is possible. 
We have tested the capabilities of CTA to detect transient VHE emission for different kinds of sources. Below we illustrate our findings with two examples:\par 

\begin{figure}[h!]
\centering
\includegraphics[width=0.8\columnwidth]{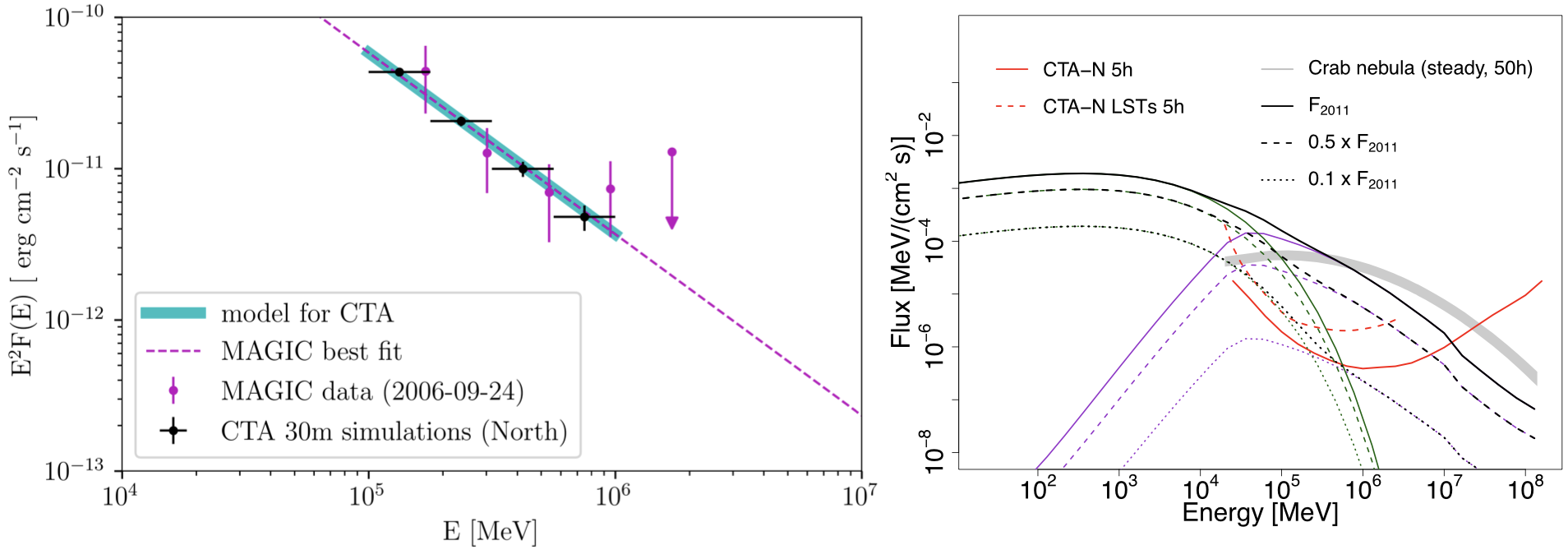}
\caption{\small{\textit{Left}: Spectral energy distribution of Cyg X-1 during a flaring episode similar to that reported in \cite{Albert2007} (magenta points). CTA will be able to detect this binary in 30 minutes of observation (black points; model in cyan). \textit{Right}: simulation of different flaring models (black lines), result of the synchrotron (green) and IC (purple) contributions. The red lines correspond to the sensitivities of CTA-North and the 4 LSTs of CTA-North (for 5 h). The steady Crab spectrum is plotted for comparison (gray shaded area). Figures from the CTA consortium paper on Galactic transients (in prep.)}} \label{fig:GalacticTransients}
\end{figure}

 \emph{Microquasars}: these binary systems are composed of a compact object (either NS or BH) accreting matter from a companion star. They are characterized by the presence of an accretion disk  and can generate collimated jets of plasma, which are normally active during certain phases (so called hard state). In a reduced number of systems the jets can be persistent as in the case of SS433, the only known multi-TeV microquasar system detected by HAWC \cite{Abeysekara2018} as an extended source up to 20 TeV. The emission is due to the interaction between the jet and the surrounding nebula, with no detection of the central binary. Emission from this binary has not yet been detected by IACTs \cite{Ahnen2018}. Simulations prove that the northern and southern CTA arrays will be able to detect SS433 and its lobes with high significance, unveiling from the first time the TeV counterpart of the central binary and covering the unexplored range between \textit{Fermi}-LAT \cite{Fermi2020} and HAWC. We have also tested whether CTA could detect transient emission from microquasars. For that purpose, we extrapolated the spectra of the hint of transient VHE gamma-ray emission observed by the MAGIC telescopes in Cyg X-1 during the 2006 flare \cite{Albert2007}. Our simulations indicate that CTA will detect Cyg X-1 during a flaring state in only 0.5 h (Fig.~\ref{fig:GalacticTransients}).
 
 
 \emph{Flares from PWNe}: PWN systems are bubbles of relativistic plasma powered by 
 a central highly magnetized rotating NS (pulsar). The best known PWN is the Crab Nebula, considered as a VHE standard candle. Although most PWNe are observed as steady sources, the discovery of rapid HE flares from the Crab Nebula \cite{Abdo2011} revealed that they are also transient sources. These flares can not be easily explained with standard PWNe theories, needing a second component to account for the variable peak emission. However, the origin of this second component is still uncertain and the gamma-ray emission seems to be inconsistent with traditional synchrotron or IC scenarios \cite{mestre2020}. CTA, with its wide energy range coverage, will help to understand the origin of these flares: the low-energy threshold will allow sampling of the \textit{Fermi}-LAT spectral shape (synchrotron tail), while the TeV regime will allow us to explore the inverse Compton (IC) component which might arise via the up-scattering of the MeV flares (see Fig.~\ref{fig:GalacticTransients}, right).

As shown, CTA will be able to detect microquasars and flares from PWNe at VHE. Additionally, its unique capabilities will likely result in the detection of a large variety of Galactic transients such as novae, flares from known gamma-ray/X-ray binaries and, on longer timescales, emission from tMSPs during accretion phase. Finally, CTA will play an important role in identifying the nature of new transients by performing follow-up observations of wide-field-of-view instruments.

\chapter{Core Collapse Supernovae}
The catastrophic explosion of a massive star at the end of its life is accompanied by the release of a tremendous amount of energy.  The energy is mainly released in the form of kinetic energy, resulting in a fast-moving shockwave that expands outwards, accompanied by emission over the entire EM spectrum, neutrinos and heavy nuclei~\citep{woosley2002}. 
Supernovae are believed to play a crucial role in the cosmic cycle of matter, star formation and galaxy formation, and in the production of cosmic rays and nuclei~\citep{branch2017}. They have been suggested as good PeVatron candidates in the first $\sim$ year after the supernova (SN) explosion~\citep{marcowith2018, Schure2013}, motivating their study in the search for the origin of Galactic cosmic rays. This is of special importance, given that the \textit{supernova remnant hypothesis} is still facing major issues~\citep[see e.g.][for pedagogical reviews]{tatischeff2018, blasi2019, gabici2019}, such as the difficulty to accelerate CRs up to the PeV range~\citep{bell2013,cristofari_pev}. 
The strong shock resulting from the explosion of the massive star has been shown to potentially be a very efficient particle accelerator. The shock expanding forward is typically 1) non-relativistic (or trans-relativistic), 2) strong (i.e., of compression factor $\gtrsim 4$), 3) accelerating particles via the first--order Fermi mechanism, and 4) expanding for several tens of kyr as a \textit{supernova remnant}. 

The accelerated particles, interacting with the surrounding interstellar matter, through pion production, should lead to a gamma-ray signal, potentially detectable with future instruments optimized in the gamma-ray domain. Observations in the VHE range will especially help address some of the following fundamental questions: What role do core-collapse supernovae (CCSNe) play in the production of CRs? What is the slope of the spectrum of accelerated particles? What is the efficiency of the acceleration process, and what are the timescales involved? What is the maximum energy of particles accelerated at CCSNe? What are the mechanisms governing the acceleration of particles at the highest energies? 

Particle acceleration is expected to be especially efficient in CCSNe, from progenitors more massive than $\gtrsim 8 M_{\odot}$. CCSNe are the most frequent SNe, accounting in our Galaxy for 2/3 of SNe~\citep{smartt2009}. In these explosions, the strong shock where particle acceleration is taking place is expanding in the dense wind produced by the late activity sequence of the progenitor. Under the assumption that a fraction of the ram pressure of the shock is converted into accelerated particles, an estimate of the gamma-ray emission from accelerated hadrons interacting with the circumstellar medium shows that a signal detectable by a next generation instrument like CTA is expected for CCSNe typically within  $\lesssim 10$ Mpc ~\citep{drury1994,tatischeff2009}.

However, a wide variety of CCSNe exist, with large variations in the properties of the progenitor star (i.e., the mass-loss rates, wind speeds), and important variations in the time evolution of the shock velocity and radius.
These variations substantially affect the gamma-ray signal. The different types of CCSNe (IIP, IIL, IIb, IIn, etc.) are thus not all equivalent in terms of production of gamma rays. Moreover, the gamma-ray signal is strongly affected by at least one mechanism: the two-photon annihilation process, in which gamma rays interact with lower energy photons emitted from the SN photosphere to produce electron-positron pairs. These mechanisms have been shown to potentially degrade the gamma-ray signal by more than 10 orders of magnitude in the first days after the explosion of the SN, in this case significantly reducing the chances of detection in the gamma-ray range~\citep{aharonian2008,murase2014,wang2019,cristofari2020}.

In this context, the unprecedented performance of CTA, above a few tens of GeV, will be a considerable asset for the study of CCSNe in the first days/months - possibly years - after the explosion, and coupled to theoretical studies, will help improve our understanding of particle acceleration in CCSNe. 




\cleardoublepage



\bibliographystyle{apsrev}
\bibliography{cta_mma}


\glsaddall
\setglossarystyle{list}

\end{document}